\documentclass{emulateapj} 
\usepackage{apjfonts}
\usepackage{epsfig}
\usepackage{xspace}
\usepackage{natbib}
\usepackage{hhline}
\usepackage{amsmath}
\usepackage{multirow}
\usepackage{dcolumn}
\usepackage{verbatim}

\newcolumntype{.}{D{.}{.}{-1}}

\newcommand{\chandra}{{\it Chandra}\xspace}
\newcommand{\xmm}{{\it XMM-Newton}\xspace}
\newcommand{\champlane}{ChaMPlane\xspace}

\newcommand{\wavdetect}{{\it wavdetect}\xspace}

\newcommand{\ipname}{{CXOPS J180354.3-300005}\xspace}

\newcommand{\sS}[1]{\mbox{$\rm{}^{#1}$}}
\newcommand{\Ss}[1]{\mbox{$\rm{}_{#1}$}}

\newcommand{\nH}{\mbox{$N_{\mbox{\scriptsize H}}$}\xspace}
\newcommand{\nHt}{\mbox{$N_{\mbox{\scriptsize H22}}$}\xspace}

\newcommand{\Bx}{\mbox{$B_X$}\xspace}

\newcommand{\Deg}{\mbox{$^\circ$}\xspace}

\newcommand{\lcgs}{\mbox{ergs s\sS{-1}}\xspace}
\newcommand{\fcgs}{\mbox{ergs cm\sS{-2} s\sS{-1}}\xspace}

\begin{document}

\title{\chandra Discovery of an Intermediate Polar in Baade's Window}


\author{
JaeSub Hong\altaffilmark{1*},
Maureen van den Berg\altaffilmark{1},
Silas Laycock\altaffilmark{2},
Jonathan E. Grindlay\altaffilmark{1},
and 
Ping Zhao\altaffilmark{1}
}
\altaffiltext{*}{Send requests to J. Hong at jaesub@head.cfa.harvard.edu}
\altaffiltext{1}{Harvard-Smithsonian Center for Astrophysics, 
60 Garden St., Cambridge, MA 02138 }
\altaffiltext{2}{Gemini Observatory, 670 N. A'ohoku Place, Hilo, HI 96720}

\begin{abstract} 

We have discovered an intermediate polar (IP) in the 100 ks \chandra
observation of Baade's Window (BW), a low extinction region at about 4\Deg
south of the Galactic Center.  The source exhibits large X-ray modulations
at a period of 1028.4 s in the 0.3 -- 8 keV band.  The X-ray spectral 
fit with a power law model shows the integrated spectrum is
intrinsically hard (photon index, $\Gamma$ =
0.44 $\pm$ 0.05) and moderately absorbed ($\nH = 1.5 \pm 1.0 \times
10^{21}$ cm\sS{-2}).  The relatively poor statistics only allows for
a mild constraint on the presence of an iron emission line
(equivalent width = $0.5 \pm 0.3$ keV at 6.7 keV).  Quantile analysis reveals
that the modulations in the X-ray flux strongly correlate with spectral
changes that are dominated by varying internal absorption. The X-ray
spectrum of the source is heavily absorbed ($\nH > 10^{22}$ cm\sS{-2})
during the faint phases, while the absorption is consistent with the
field value ($\sim 10^{21}$ cm\sS{-2}) during the bright phases.
These X-ray properties are typical signatures of IPs.  Images taken with
the IMACS camera on the Magellan 6.5m telescope show a faint ($V \sim
22$), relatively blue object ($B_0-V_0\gtrsim 0.05$) within the 2$\sigma$
error circle of the \chandra source, which is a good candidate for being
the optical counterpart.  If we assume a nominal range of absolute $V$
magnitude for a cataclysmic variable ($M_V \sim 5.5-10.5$) and the
known reddening in the region ($A_V =1.4$ at $> 3$ kpc), the source
would likely be at a distance of 2--10 kpc and not in the local solar
neighborhood. The corresponding average X-ray
luminosity would be $6 \times 10^{31} - 10^{33}$ \lcgs in the 2--8
keV band.  Assuming  the space density of IPs  follows the stellar
distribution, which is highly concentrated in the Galactic Bulge,
the source is probably a relatively bright IP ($\sim 10^{33}$ \lcgs if
it is at 8 kpc) belonging to the Galactic Bulge X-ray population, 
the majority of which is now believed to be magnetic
cataclysmic variables.

\end{abstract}

\section{Introduction}
\setcounter{footnote}{2}

Intermediate Polars (IPs) are a type of magnetic cataclysmic variable (CV),
where the magnetic field of the accretor, a white dwarf (WD), disrupts
the inner portion of the accretion disk and channels the
accretion flow into the magnetic poles of the WD \citep{Patterson98}. 
As the WD spins,
this channeling gives rise to a pulsation, a tell sign of an IP, which
is typically found in the period range of $\sim$ 100 -- 1000s.
The magnetic field on the surface of the WD in IPs is typically
about $10^{6 - 7}$G, and the ratio of the spin to orbital period of
IPs is found in a wide range from 0.01 to $<$1.  
Some IPs are on the evolutionary 
path to their cousins, the polars, where the orbital and WD spin periods are
usually locked under the strong magnetic fields that convert the whole
accretion flow into a stream \citep{Norton04}.  Magnetic CVs provide
unique astrophysical laboratories for studying physics in extreme
conditions, stellar and binary evolution, which is important to make a
census of Galactic population and understand their evolution.  There are
about 33 confirmed  and 72 candidate IPs today, and
the full catalogue is found in \citet{Ritter03}\footnote{See
the on-line catalogue at \url{http://physics.open.ac.uk/RKcat/} and
\url{http://asd.gsfc.nasa.gov/Koji.Mukai/iphome/iphome.html}
for the latest update.}.

The 72 IP candidates contain seven periodic X-ray sources at the
Galactic Center (GC), which are part of $\sim$ 2300 X-ray sources
discovered in the deep \chandra observations of
the Sgr A* field \citep{Muno03a}.  The exact nature of the majority of
the GC X-ray population still remains elusive: direct identification
at other wavelength is difficult due to high obscuration by dust and
source confusion due to high star density. Therefore, the discovery of
these IP candidates is particularly interesting since it supports the
idea of magnetic CVs as the leading candidates for the
majority of the GC X-ray sources \citep{Muno04,Laycock05}.  If true,
this also implies the GC X-ray population is very old.

In order to explore the X-ray population in the Galactic
Bulge (GB) without obscuration by the intervening dust, we have observed
three low extinction Windows within 4\Deg of the GC with the \chandra X-ray
observatory \citep{Berg06,Berg09,Hong09}. The initial population study
on the X-ray sources in these Windows and four GB fields has revealed
that the GB X-ray sources extend out to $>1.4\Deg$ from the GC with
a projected source density that follows a $1/\theta$ relation, where
$\theta$ is the angular offset from the GC
\citep{Hong09}.  In addition, the similarity of the X-ray spectra of the
hard X-ray sources in these fields -- an intrinsically hard continuum
with the presence of an iron emission line -- indicates a single class
of sources, likely magnetic CVs, could make up both the GC and GB X-ray
populations.

As part of our efforts to identify the nature of these sources, we have searched
for periodic modulations in the X-ray emission of the bright X-ray sources
(net counts $\ge$ 250 in the 0.3--8 keV range) in the Window fields
using the Lomb-Scargle algorithm \citep{Scargle82}.  As a result, we
have discovered an IP in Baade's Window (BW) based on
a strong pulsation correlated with the X-ray spectral variation
(\S\ref{s:obs}).
We explore the properties of this X-ray source (\S\ref{s:xray}) and the
potential optical counterpart (\S\ref{s:opt}) that lead to its
identification as an IP. We discuss the possibility that the system belongs
to the GB X-ray population (\S\ref{s:dis}).

\section{Observation and Timing Analysis} \label{s:obs}

We have observed BW on 2003 July 9 (Obs.~ID 3780), Stanek's Window (SW)
on 2004 February 14/15 (Obs.~ID 4547 and 5303), and
Limiting Window (LW) on 2005 August 19/22 and October 25 (Obs.~ID
5934, 6362 and 6365) with the \chandra ACIS-I instrument 
\citep{Berg06,Berg09,Hong09}.  The data were analyzed as a part of our survey
program, the \chandra Multi-wavelength Plane (\champlane) survey,
which is designed to constrain the Galactic
population of low luminosity
accretion sources, CVs in particular \citep{Grindlay05}. In summary, we
search for X-ray point sources using a wavelet
algorithm (\wavdetect, \citealt{Freeman02}) and perform aperture
photometry to extract the basic X-ray properties.
The details of the analysis procedures are described in \citet{Hong05,Hong09}.

In order to find periodic X-ray modulations in the light curves of these
X-ray sources, we have performed a Lomb-Scargle periodicity search on the bright
X-ray sources (net counts $\ge 250$ in 0.3--8.0 keV) discovered in the
100 ks observation of the Window fields - BW (Obs.~ID 3780), Stanek Window
(Obs.~ID 4547 and 5303) and Limiting Window (Obs.~ID 5934, 6362 and 6365)
\citep{Hong09}.  The events are selected by the good time intervals
(GTIs), where the background fluctuation is less than 3$\sigma$ above
the mean level \citep{Hong05}.
The arrival time of each photon is bary-center
corrected using the CIAO tool {\tt
axbary}\footnote{\url{http://cxc.harvard.edu}}. For each source,
we generate the light curves in time bins of four multiples (1,
4, 8 and 16) of the CCD integration time (3.2 s). Then we apply the
Lomb-Scargle algorithm and calculate the power spectrum 
for the light curves at all four time resolutions.

\begin{figure}[t] \begin{center} 
\epsscale{1.0}
\plotone{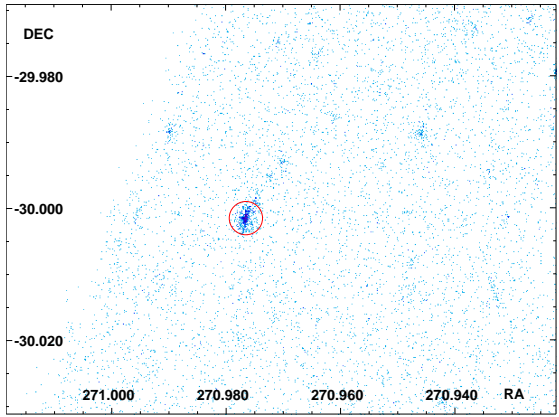} 
\end{center} 
\caption{The raw X-ray sky image around \ipname. 
The (red) circle indicates the source region used for the aperture photometry.
The morphology of the event distribution is consistent with
the expected point spread function at the source location.}
\label{f:raw} \end{figure}

We have found that one source in BW, \ipname  exhibits a clear sign
of a periodic modulation 
with greater than 99\% of the confidence level (CL).  The source
was detected with $510 \pm 24$ net counts in the broad band
(\Bx, 0.3--8 keV) after background subtraction.  Fig.~1 shows the
raw sky image around the source marked by the (red) circle of the 95\%
point spread function (PSF).  The source only appears to be extended due to
the large offset (8.5\arcmin) from the aimpoint of the instrument, but
a simulated PSF by a CIAO tool {\tt mkpsf} at the source position closely
resembles the event distribution of the source, indicating that the source
is consistent with a point source.

Fig.~2a shows the power spectrum of the source in the frequency domain
along with horizontal lines indicating 90, 95 and 99\% CLs. This power
spectrum is based on the light curve at the 12.8 s time resolution and
the results based on the other three time resolutions are very similar.
The power spectrum indicates the source exhibits pulsations at a period
of 1028.4 $\pm$ 3.8 s.  The error of the period estimate is the 1$\sigma$
equivalent spread of the peak in the power spectrum, which is calculated
by a fit to the peak with a Gaussian function in the period domain.
The observed period is somewhat close to a possible spurious period
(1000 s). Sources falling near the edges or node boundaries of the chips
can exhibit spurious modulations at a period of 707 or 1000 s
(or their harmonics) due to the dither motion of the \chandra X-ray
observatory.  However, we consider the periodic modulation found in
\ipname to be real because first, the source is not near the chip edges
or the node boundaries and second, the X-ray spectral properties of the
source strongly correlate with the flux modulation as demonstrated below.

Fig.~2b shows the folded light curves at the pulsation period in the soft
(0.3--2.0 keV), hard (2.0--8.0 keV) and broad (0.3--8.0 keV) bands.
The folded light curve of the broad band is shifted up by 1 cts ks\sS{-1}
for clarity.  The smooth lines are calculated using the
LOWESS algorithm \citep{Cleveland94} and color-coded by the phase for
a later reference in the spectral analysis (\S3). The data points are
calculated at the 10 equal-sized phase bins of width 0.1.  The net count
per bin ranges from $\sim$ 18 to 90 in the 0.3--8.0 keV range.

\begin{figure*}[t] \begin{center} 
\epsscale{1.0}
\plotone{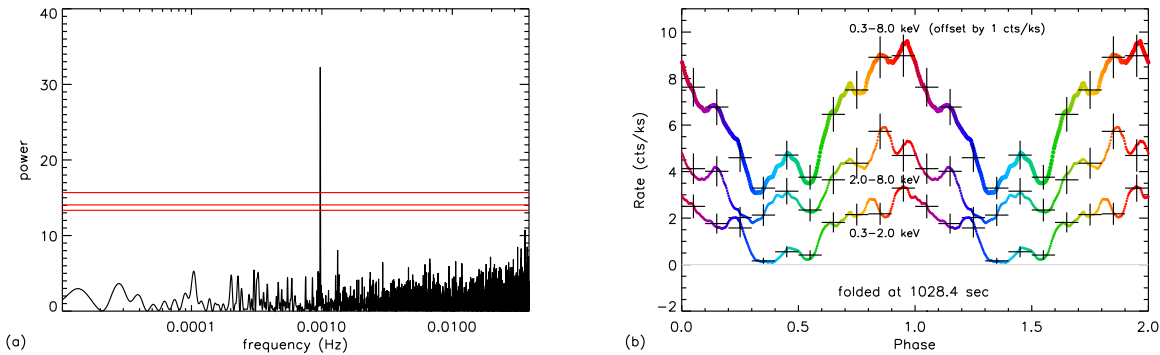} 
\end{center} 
\caption{The power spectrum (a) of \ipname in the broad band (0.3--8.0 keV)
and the folded light curves (b) in the soft (0.3--2.0 keV), hard
(2.0--8.0 keV), and broad bands (phase = 0 at JD 2452830.0).
The primary modulation is found
at 9.727 $\pm$ 0.036 $\times 10^{-4}$ Hz or 1028.4 $\pm$ 3.8 s.
The horizontal lines in (a) represent the confidence levels (CLs) of
90, 95 and 99\%. 
The smooth lines in (b) are generated by the LOWESS
alghorithm and color-coded by the phase for a later reference (Fig.~\ref{f:sa}). 
The data points are calculated in the equal phase bins (0.1). The
modulation depth is estimated to be $(90 \pm 10)\%$, $(42 \pm 10)\%$
and $(53 \pm 8)\%$ for the soft, hard, and broad bands respectively.
The folded light curves in the broad band are shifted up by 1 cts
ks\sS{-1} for clarity.}
\label{f:pds} \end{figure*}

We define the modulation depth as
$(R\Ss{max}-R\Ss{min})/(R\Ss{max}+R\Ss{min})$ where $R\Ss{max}$ and
$R\Ss{min}$ are the maximum and minimum of the fitted amplitudes
respectively.  Using a fit to the 10 data points with a sinusoidal
function, we estimate the modulation depth is $(90 \pm 10)\%$, $(42
\pm 10)\%$ and $(53 \pm 8)\%$ for the soft, hard, and broad bands
respectively.  The modulation amplitude varies with the energy band,
which we will explore in more detail using
quantile analysis in \S\ref{s:xray}.

BW was also observed with the EPIC cameras on the  \xmm observatory in
two separate pointings with $\sim$ 20--25 ks exposure each on 2002 March
11 and 2004 September 30.   The data from both observations are
publicly available.  Our source was detected as 2XMM
J180354.3-300004\footnote{\url{http://xmm.esac.esa.int/xsa/}}
at large offsets (5.7\arcmin\ and 7.6\arcmin) from the aimpoint in
both observations \citep{Watson09}.  Unfortunately, in the four datasets out of six total
(three cameras and two pointings), the source fell near a chip gap. In
both PN observations, which would have provided an interesting result
because of the superior sensitivity at high energies than the \chandra
ACIS instruments, the central part of the PSF overlapped with a chip
gap, rendering the data practically unusable.  Three MOS observations
(including one with the source marginally close
($\sim$30\arcsec) to the chip edge) appear to produce relatively {\it
clean} dataset of the source, but the poor statistics (net counts:
$\sim 60 - 80$ each in 0.3--8 keV) did not allow for any significant
detection of pulsation.  Therefore, in the following, we mainly consider
the \chandra data for the X-ray observation of the source.  See Table 1
for the overall flux estimates of the source from the \xmm observations,
which is consistent with the same from the \chandra data.


\section{X-ray Spectrum and Variation} \label{s:xray}

\begin{table*}
\caption{Spectral Model Fits and Quantile Analysis of the integrated X-ray spectrum of \ipname}
\begin{tabular*}{\textwidth}{l@{\extracolsep{\fill}}cccccccccc}
\hline\hline
	& \multicolumn{4}{c}{Spectral Fits}						& \multicolumn{3}{c}{Quantile Analysis}			& \multicolumn{2}{c}{Flux} 	\\
\cline{2-5} \cline{6-8} \cline{9-10}
Model  	& \nHt		 		& $\Gamma$ 	& EW		& $\chi^2$/DoF	&  \nHt		 		& $\Gamma$ 	& EW	& 0.5 -- 2 keV &	2 -- 8 keV 	\\
	& ($\times10^{22}$cm\sS{-2})	&		&(keV)		& 		&($\times10^{22}$cm\sS{-2})	&		& (keV)	&\multicolumn{2}{c}{($\times 10^{-13}$ \fcgs )}	\\
\hline
\multicolumn{4}{l}{\chandra} \\
PL 	& 0.15(10) 	 		&0.44(5)       	& -		& 12.6/22	& 0.28(21)			& 0.49(18)	& - 	& 0.13(1) 	&1.20(7)\\
PL+Fe 	& 0.22(10)	 		&0.53(5)       	& 0.5(3)	& 10.3/20	& 0.39(22)			& 0.63(18)	& 0.5	& 0.14(1)	&1.19(7)\\
\hline
\multicolumn{4}{l}{\xmm MOS} \\
PL  &  	 		&       	& 		& 		& $0.61_{-0.41}^{+0.64}$			& $0.60_{-0.39}^{+0.41}$	& - 	& 0.19(12) 	&1.25(43)\\
\hline
\end{tabular*}
For the \chandra data, the flux estimates are based on the spectral model
fits. The errors  are of statistical origin, based on
the photon counts using small-number statistics by \citet{Gehrels86}.
The PL+Fe model in quantile
analysis assumes a 0.5 keV EW for the iron line at 6.7 keV. 
In the \xmm data, we combine three datasets and use quantile analysys
under a simple power law. The datasets comprise the MOS1 observation in the 2002
pointing, and both the MOS1 \& MOS2 observations in 2004. The flux
estimate is based on the quantile analysis of the combined dataset,
and the errors are the quadratic sum of the statistical error and the
standard deviation of the three datasets. The reported flux 
in 0.2--12 keV is $1.9(2) \times 10^{-13}$ and $1.6(2) \times 10^{-13}$ \fcgs
for the 2002 and 2004 \xmm observations respectively \citep[
see also \url{http://xmm.esac.esa.int/xsa/}]{Watson09}.
DoF stands for degrees of freedom.
\label{t:fit}
\end{table*}

\begin{figure}[t] \begin{center} 
\epsscale{1.0}
\plotone{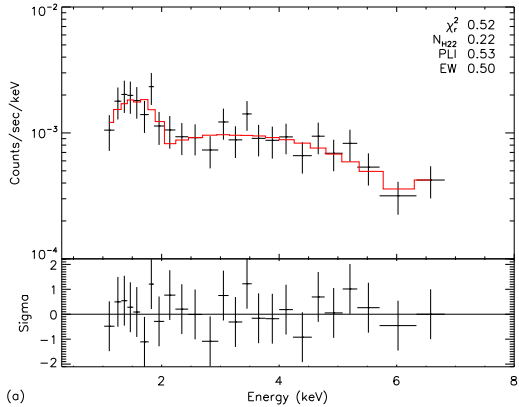} 
\end{center} 
\caption{The spectral model fit to the integrated X-ray spectrum of \ipname
under a power law plus an iron line at 6.7 keV.
See also Table 1. The spectral fit shows a hint of the iron emission line. }
\label{f:fit} \end{figure}

Table 1 summarizes the results of the spectral model fits to the
integrated X-ray spectrum of the source using a power law model with
and without an iron emission line at 6.7 keV.  The Fe He-$\alpha$ XXV
line is considered because many of the GC X-ray sources exhibit
this line more prominently than the neutral Fe line at 6.4 keV
\citep{Wang02,Muno03,Hong09}.  Both models
produce acceptable fits, and the spectrum is intrinsically 
hard ($\Gamma \sim 0.4$) and moderately absorbed ($\nHt \sim 0.2$).
Fig.~\ref{f:fit} shows the spectral model fit to the integrated spectrum
of the source using the power law plus the iron line.  There is a hint
of an emission line in the 6 -- 7 keV energy range in the spectrum but
the poor statistics does not provide a significant constraint on the
presence of the iron line. The estimated equivalent width (EW) of the
iron line at 6.7 keV is 0.5 $\pm$ 0.3 keV.  In the case of the thermal
Bremsstrahlung or thermal plasma model, we can only assign a lower
limit for the temperature ($>10$ keV with 95\% confidence level for
thermal Bremsstrahlung or $>7.0$ keV for MeKaL) and we cannot confirm
this apparent high temperature is intrinsic due to low statistics and
lack of the spectral information at high energies ($>10$ keV).

The total column density in the direction of \ipname is estimated to
be \nHt = 0.25(8) 
by \citet{Marshall06} or 0.25(5) by \citet{Sumi04}, using $\nHt =
0.179\ A_V$ \citep{Predehl95}.  The former estimates are given in a
set of distances starting from 2.75 kpc for the BW field and the latter
estimates are given at a finer angular resolution ($\sim$ 0.5\arcmin\ --
2.0\arcmin) than the former ($\sim$ 8.0\arcmin).  There are also other
estimates for the extinction in the BW
such as $\sim$ 0.34 by \citet{Schlegel98} or $\sim$ 0.41 by
\citet{Drimmel03}, but the underlying models of these estimates are not
accurate in this region.

In order to explore the spectral variation as a function of the modulation
phase, we use quantile analysis. Quantile analysis is a bias-free spectral
classification method, suitable for faint sources or for investigating
subsets of the data that do not allow for spectral model fits due to 
low statistics \citep{Hong04}.  Unlike the conventional hardness ratio,
which is subject to the spectral bias intrinsic to the choice of the
sub-energy ranges, quantile analysis provides a reliable spectral
measure of sources even with counts as low as $\sim$ 10 because it takes 
full advantange of the given statistics without sub dividing the energy 
range \citep[e.g.][]{Berg06}.

\begin{figure*}[t] \begin{center} 
\epsscale{1.0}
\plotone{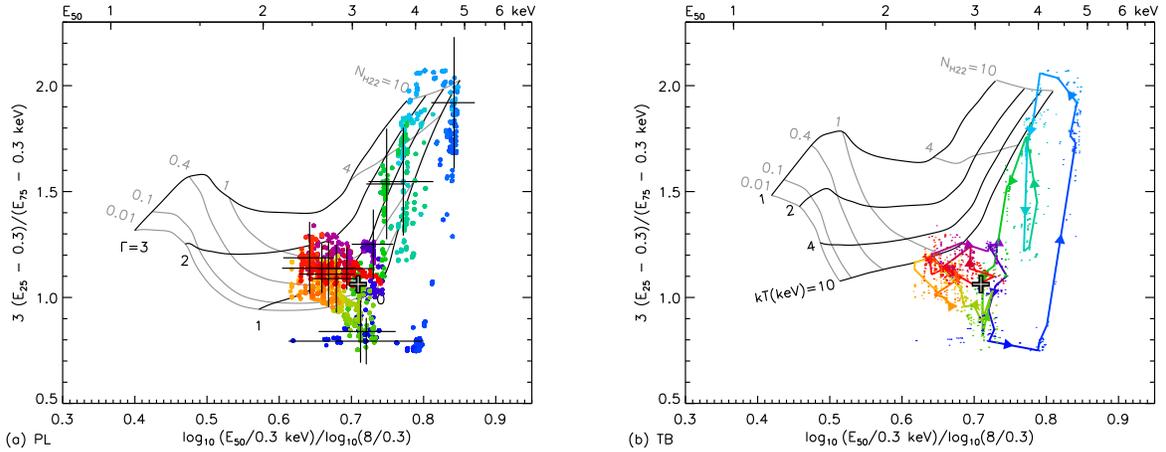} 
\end{center} 
\caption{ The quantile diagram
of \ipname as a function of modulation
phase, color-coded to match the phase in Fig.~\ref{f:pds}b (see the electronic version).
Both panels show the same data and the 10 binned data points are shown
in (a).
The grid in (a) is for simple power law models with
$\Gamma$ = 0, 1, 2 \& 3 and \nHt = 0.01, 0.1, 0.4 , 1, 4 \& 10
and in (b) for thermal Bremsstrahlung model with $kT$ =
1, 2, 4, 10 keV and the same \nHt.
The arrows in (b) show the temporal orientation of the spectral change. 
There is a strong
correlation between the quantiles and the modulation phase.  
The energy quantile $E_p$ corresponds to the energy
below which $p\%$ of the counts are detected in the 0.3--8.0 keV band. 
The top-axis of the
quantile diagram is labeled by the median energy ($E_{50}$).}
\label{f:sa} \end{figure*}

We calculate the quantile values of the
X-ray spectra of the source  as a function of modulation phase using a
sliding phase window of a fixed size (0.1).  Fig.~\ref{f:sa} shows such
a quantile diagram.  The energy quantile $E_p$ corresponds to the energy
below which $p\%$ of the counts are detected in the energy
range (0.3--8.0 keV in Fig.~\ref{f:sa}). For instance, $E_{50}$  means
the median energy, and $E_{25}$ and $E_{75}$ are two quartiles.  Note
the definition of the $x$-axis in Fig.~\ref{f:sa} is different from
the one suggested by \citet{Hong04}.  The new definition
is $\log_{10}(E_{50}/E\Ss{lo}) / \log_{10}(E\Ss{hi}/E\Ss{lo})$ in
general, where $E\Ss{lo}$ and $E\Ss{hi}$ are the lower and upper bound
of the energy range respectively ($E\Ss{lo}$=0.3 and  $E\Ss{hi}$=8.0
keV in Fig.~\ref{f:sa}).  We believe the new definition is
more reflective of the instrument response and 
the confined range (0--1) allows for statistically more uniform 
response throughout the phase space of quantile diagram
\citep{Hong09b}.

For easy interpretation, the quantile diagram is overlaid with power
law and thermal Bremsstrahlung model grids.  The filled circles are
color-coded to match the phase of the smooth line in the folded light
curve in Fig.~3b for easy comparison and the data points with the
error bars are from the same 10 data points in the folded light curve.
Table 1 also shows the spectral parameters estimated by quantile analysis
of the integrated spectrum (marked by
a hollow cross in Fig.~\ref{f:sa}) and the results
are consistent with the spectral fit.

The quantile diagram reveals dramatic spectral changes as a function of
pulsation phase, illustrated by a large loop formed counter-clockwise as
the X-ray flux modulates in a full cycle (see the track in Fig.~\ref{f:sa}b).
The spectral variation is expected from the energy dependence of the modulation depth
seen in Fig.~\ref{f:pds}. In the quantile diagram, we can also
see that the spectral changes are dominated by varying extinction.  
During the bright phases,
the spectrum is relatively unabsorbed with \nHt $\lesssim 1$, less than
or consistent with the field extinction (e.g. the yellow and red points in
Fig.~\ref{f:pds}b \& \ref{f:sa}b), but during the faint phases, especially
around phases 0.3 -- 0.6 (e.g. the light blue points), the spectrum is
heavily absorbed with \nHt $>5$.  Since the field extinction is very low,
the additional increase in extinction must be intrinsic to the source.
Note the long axis of the data loop in the quantile diagram is largely
parallel to the direction related to changes in absorption
for both spectral models as shown in Fig.~\ref{f:sa}. 
In fact, this is true in general regardless of the specral
models.

Table~\ref{t:fit} also shows the results of the \xmm data for comparison,
for which we combine three relatively clean datasets (the MOS1 observation
in 2002, both MOS1 and MOS2 in 2004).  The overall spectral properties and
the flux estimates of the \xmm data based on the quantile analysis are
consistent with those derived from the \chandra data.

\section{Optical counterpart} \label{s:opt}

\begin{figure*}[t] \begin{center} 
\epsscale{1.0}
\plotone{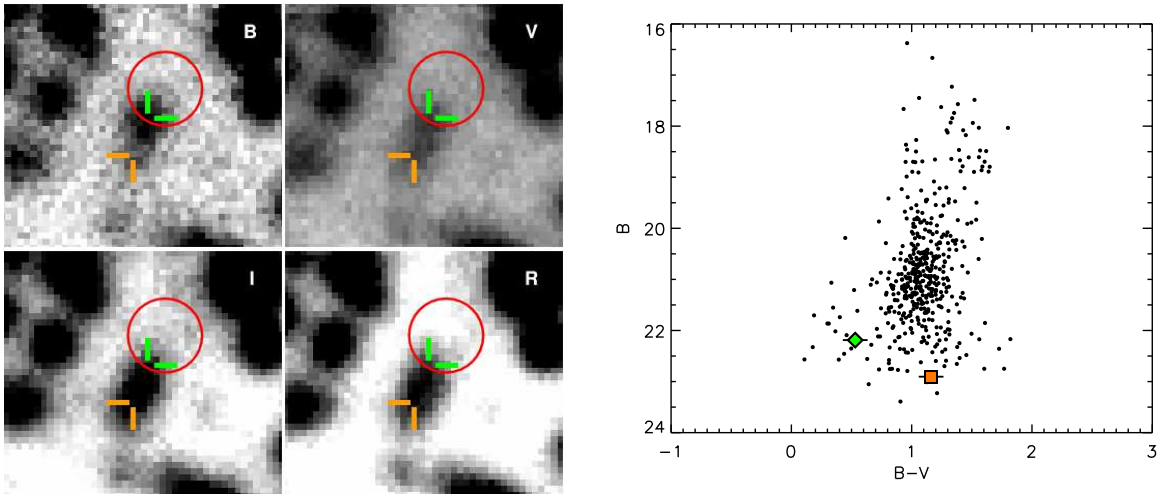} 
\end{center} 
\caption{The finding chart in the $B, V, I$ and $R$ band images (left)
and the color-magnitude diagram (CMD) in $B$ vs.~$B-V$ (right).
In the finding chart the (red) circle centered at the X-ray source
(radius: $2\sigma$ of the composite errors, 0.66\arcsec) indicates the
search region for counterparts. The candidate counterpart and
the neighbor star are marked by green and orange tick marks
respectively (white tick marks for the print edition).  The CMD
compares the two marked sources with the field stars within 30\arcsec.
The potential counterpart (green diamond) is mildly bluer than the
most field stars. }
\label{f:opt} \end{figure*}

On 2007 May 8, we observed the three Window fields
with the Inamori Magellan Areal Camera and Spectrograph
(IMACS) on the 6.5 m Magellan (Baade) telescope at Las
Campanas, Chile. Under good conditions (seeing FWHM $\sim 0.5\arcsec$, clear sky) we
obtained a dithered set of 5 pointings in the f/4 configuration (15\arcmin\ field,
0.2\arcsec/pixel) to cover an $18\arcmin \times 18\arcmin$ region of BW. This
provided a total exposure time of 500 s in each of Bessell-$B$, $V$, $R$, \&
CTIO-$I$ filters over the \chandra field.

We processed the images using standard IRAF tasks, and calibrated the
astrometry using the 2MASS catalogue as a reference. The astrometric
residuals on each CCD frame were $\sim$ 0.2\arcsec. We reprojected and stacked the
images using the SWARP\footnote{\url{http://terapix.iap.fr}} utility. All frames
were normalized to ADU/second units and combined using weight-maps
constructed from flat-fields
and bad pixel masks. The initial source search and photometry
were performed on the stacked images using SExtractor \citep{Bertin96}.

After boresighting the initial IMACS source list to the
\chandra sources \citep{Zhao05}, and applying offsets of $\Delta$RA~= 0.088(51)\arcsec,
 $\Delta$DEC = 0.498(50)\arcsec\ to the X-ray coordinates, we found a potential
counterpart in the $B$ and $V$ band images within the search area
of the \chandra source. In Fig.~\ref{f:opt}a, the source search area
is marked by the (red) circle, the radius of which is the 2$\sigma$
quadratic sum of the boresight error and the positional errors
of the X-ray and optical sources \citep{Zhao05}. The potential counterpart
(indicated with green tick marks in the figure) is not completely resolved
with a nearby star (orange). Therefore, we generate a PSF model using a
few bright isolated sources within $\sim 30\arcsec$, and perform  PSF
fitting to the two sources in the $B$ band image (where the candidate
counterpart is brightest and its neighbor is faintest)
using the {\tt DAOPHOT} package {\tt ALLSTAR} in IRAF in order to derive the position and
magnitude of the source. The position of the optical source is
(RA, DEC) J2000 = (18:03:54.45, $-$30:00:06.3). For the $V$, $I$ and $R$
band images, we performed PSF fitting with
the fixed positions given by the results of the PSF fitting in the
$B$ band image. The instrumental $B$, $V$ and $I$ magnitudes are calibrated
against OGLE-II \citep{Udalski02} stars in the field by first order
polynomial fits.

Fig.~\ref{f:opt}b shows a calibrated color-magnitude diagram (CMD)
for stars within $\sim 30\arcsec$ of the potential counterpart (green
diamond) and the neighbor (orange square). The CMD shows the potential
counterpart is bluer than the majority of field stars. The calibrated
apparent magnitude of the optical counterpart is $B$ = 22.2(1), $V$ =
21.7(1) and $I$ = 20.6(1) where the errors are the output of the
{\tt DAOPHOT}.  Taking the interstellar extinction to BW to be $A_V$ =
0--1.4, $E(B-V)$ = 0--0.45 leads to a de-reddened color index $(B-V)_0$
= 0.05--0.5 for the optical counterpart. This value is in the typical
range for cataclysmic variables,
see for example \citet{Allen77}, table 17.9.  The number of blue sources
(24 sources with $B-V< 0.7$) falling in the error circle by chance is
about 0.03.

We estimate the ratio of the unabsorbed X-ray to optical
flux, $\log(F_X/F_V)$ = 0.63--1.12 for $A_V$ = 0.0--1.4, where
$F_X$ is in the 2--8 keV range. The blue color of the optical
source and the relatively high value  of $\log(F_X/F_V)$ 
(typically $<-2.5$ for normal stars)
are consistent with the accretion nature of the X-ray emission
from the source, indicating that the optical source is a good
candidate for being the counterpart.

\section{Discussion} \label{s:dis}

The lack of a bright optical counterpart ($V\lesssim 14$) rules out a high mass
X-ray binary (HMXB) and hence a neutron star pulsar for the system.
The relatively hard spectrum rules out a quiescent low mass X-ray binary
or a coronal nature of the X-ray emission.  In fact, the X-ray properties
of the source show classic signatures of IPs such as the intrinsically
hard X-ray spectrum with, although marginal, a possible iron emission
line and the X-ray pulsation correlated with the internal absorption
at a period of 1028 s.  The estimated luminosity of the system (see
below) is also consistent with IPs.  A likely scenario for modulating
internal absorption at the spin period is that in a bright phase we have
a relatively uninterrupted view of the emission region on the surface
of the WD whereas in a faint phase we are looking at the emission region
through the channelled accretion stream, which attenuates the X-rays.

In the following, we estimate the likely distance of the source 
to see if the system could belong to the GB population,
the majority of which are believed to be IPs.  If we assume a typical
range of absolute $V$ magnitude for  a CV
($M_V$ $\sim$ 5.5--10.5) \citep{Patterson98} and the 
known reddening in the region ($A_V =1.4$ at $> 3$ kpc)
\citep{Marshall06}, the source distance should be in the range of
2--10 kpc for $V\sim 22$, suggesting that the source is non-local.
Note if this optical source is not the counterpart of the X-ray source
($V \gtrsim 22$), the distance estimate becomes the lower bound,
which reinforces the argument that the X-ray source is non-local.
The corresponding average X-ray
luminosity would be $6 \times 10^{31} - 10^{33}$ \lcgs in the 2--8 keV
range.  According to the population synthesis model for the magnetic
CVs in the GC region by \citet{Ruiter06,Heinke08}, this luminosity range
is acceptable for magnetic CVs.

\begin{figure}[t] \begin{center} 
\epsscale{1.0}
\plotone{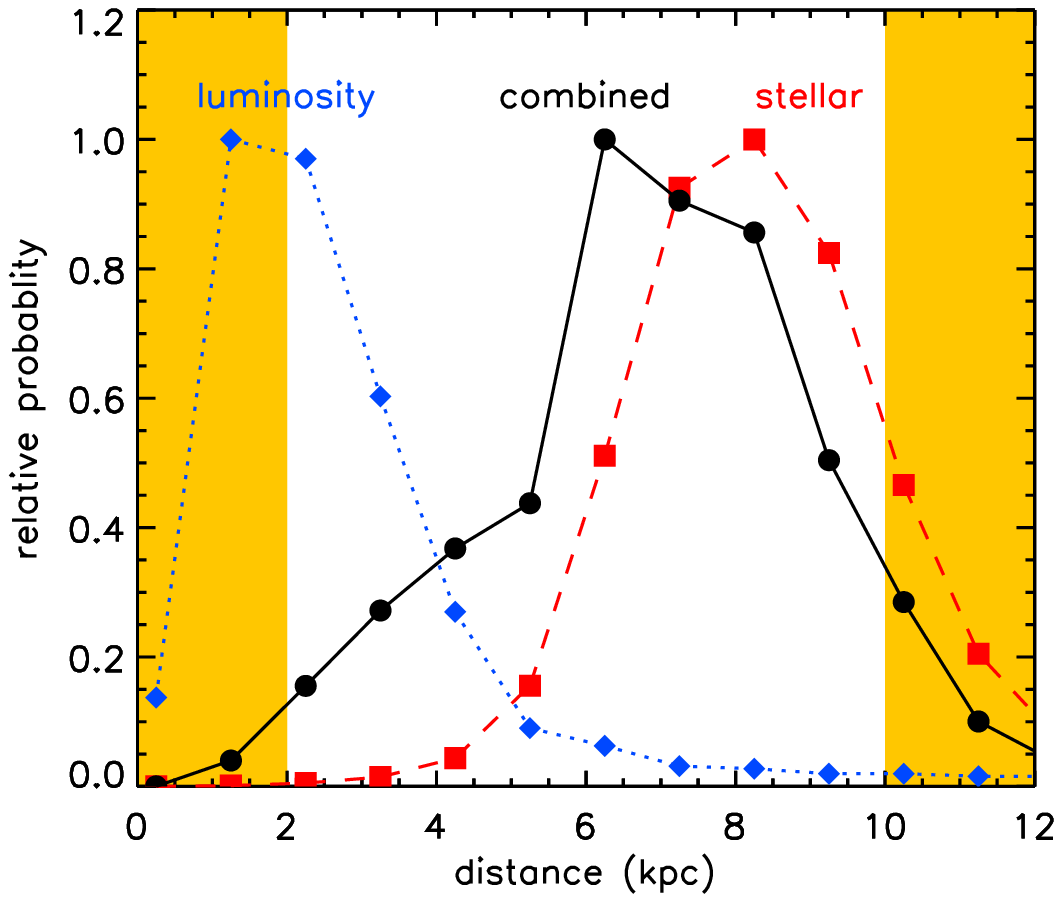} 
\end{center} 
\caption{The combined probability distribution of the source distance
(black circles) from the stellar distribution (red squares, model A in
\citet{Hong09}) and the distance distribution based on the X-ray
luminosity (blue diamonds, the standard model in \cite{Ruiter06}) for
the given flux of the source ($1.2 \times 10^{-13}$ \fcgs). The
unshaded region represents the acceptable distance for the source
based on the $V$ magnitude of the potential counterpart. Each
distribution is rebinned in 0.5 kpc intervals.}
\label{f:dp} \end{figure}

We can take this a step further.
Assuming the space density of IPs follows the stellar density, in BW 
we expect IPs are more likely found in the GB at a distance of 6--10 kpc.
For instance, if we use the stellar model (model A) in \citet{Hong09},
shown as (red) squares in Fig.~\ref{f:dp},
the probability that a given IP in the field is at 6--10 kpc is about 80\% when there is
no luminosity constraint.
We can convert the luminosity distribution of IPs in
\citet{Ruiter06} (their standard model) to a probable distance
distribution for the given flux of the source
($1.2 \times 10^{-13}$ \fcgs in 2--8 keV), which is shown as (blue) diamonds
in Fig.~\ref{f:dp}.
Now if we combine these two distance-dependent constraints,  we can estimate
the probability distribution of the source distance for the given X-ray flux
(black circles). 
The result indicates there is about a 70\% chance that the source is 
at a distance of 6--10 kpc among the possible range of 2--10 kpc set by 
the $V$ magnitude of the potential counterpart (64\% chance without
any constraint from the potential counterpart). 
This estimate mildly favors the source being a bright IP in the GB
X-ray population rather than an IP outside of the GB.
Note there are large uncertainties in the assumptions of this estimate.
For instance, the statistics in the model of the X-ray luminosity distribution 
at $> 10^{32}$ \lcgs is relatively poor (see Fig.~1 in \citet{Ruiter06}).

If indeed there are numerous IPs among the hard X-ray sources found in
this region, long term X-ray monitoring can lead to a 
direct identification of many sources through detection of pulsations. 
For instance, for a 1 Ms exposure of the BW, we expect to detect $>$
500 counts from IPs with 10\sS{32} \lcgs in the GB at 8 kpc, which
allows for a direct detection of modulations if present. 
A crude scaling using $N(>S)\sim S^{-1.5}$ predicts we should be able
to identify $\sim$ 30 sources or more in BW from such an exposure, and at
10\sS{32} \lcgs or below, we start to see the major part of the luminosity
distribution of the magnetic CVs according to
\citet{Ruiter06,Heinke08}.  In addition, low extinction regions such
as the Window fields are better suited for this approach than
high extinction fields such as Sgr A* since the dust in the high
extinction fields would attenuate the apparent X-ray modulation or spectral
changes dramatically, which would make the pulsation
very hard to detect.   For example, the spectral variation seen
in Fig.~\ref{f:sa} would not have been easily identifiable with the
substantial external extinction ($\nHt \sim 6$, a nominal value for
the GC \citep{Baganoff03}) if the source was located in the GC in the
Sgr A* field.

We thank A.~Ruiter for providing the X-ray luminosity distribution of IPs.
This work is supported in part by NASA/\chandra grants GO6-7088X, GO7-8090X
and GO8-9093X. 



\end{document}